\documentclass[12pt,thmsa]{article}%
\usepackage{amssymb}
\usepackage{sw20elba}
\usepackage{amsmath}
\usepackage{amsfonts}
\usepackage{graphicx}%
\providecommand{\U}[1]{\protect\rule{.1in}{.1in}}

\begin{document}

\author{J.J. Hamlin, S. Deemyad$^{\ast}$, and J. S. Schilling\\\textit{Department of Physics, Washington University}\\\textit{C. B. 1105, One Brookings Dr., St. Louis, Missouri 63130}%
\vspace*{0.6cm}
\and M.K. Jacobsen, R.S. Kumar, and A.L. Cornelius\\\textit{HiPSEC and Department of Physics, University of Nevada}\\\textit{Las Vegas, Nevada 89154}
\and G. Cao\\\textit{Department of Physics \& Astronomy, University of Kentucky}\\\textit{Lexington, Kentucky 40506}
\and J.J. Neumeier\\\textit{Department of Physics, P.O. Box 173840}\\\textit{Montana State University, Bozeman, Montana 59717}}
\title{Studies on the weak itinerant ferromagnet SrRuO$_{3}$ under high pressure to
34 GPa}
\date{March 16, 2007}
\maketitle

\begin{abstract}
The dependence of the Curie temperature $T_{Curie}$ on nearly hydrostatic
pressure has been determined to 17.2 GPa for the weak itinerant ferromagnetic
SrRuO$_{3}$ in both polycrystalline and single-crystalline form. $T_{Curie}$
is found to decrease under pressure from 162 K to 42.7 K at 17.2 GPa in nearly
linear fashion at the rate $dT_{Curie}/dP\simeq-6.8$ K/GPa. No
superconductivity was found above 4 K in the pressure range 17 to 34 GPa.
Room-temperature X-ray diffraction studies to 25.2 GPa reveal no structural
phase transition but indicate that the average Ru-O-Ru bond angle $\beta$
passes through a minimum near 15 GPa. The bulk modulus and its pressure
derivative were determined to be $B_{o}=192(3)$ GPa and $B_{o}^{\prime
}=5.0(3),$ respectively. Parallel ac susceptibility studies on polycrystalline
CaRuO$_{3}$ at 6 and 8 GPa prsesure found no evidence for either
ferromagnetism or superconductivity above 4 K.

\vspace{1cm}\noindent$^{\ast}$Present address: \ \textit{Harvard University,
Lyman Laboratory of Physics, Cambridge, Massachusetts 02138}

\end{abstract}

\newpage

\section{Introduction}

In all 4$d$ elemental metals, including Ru, the degree of overlap between
neighboring 4$d$ orbitals is far too great to permit the formation of either
local-moment or itinerant magnetism \cite{schilling1}. One can estimate,
however, that if one were to increase the interatomic separations in Ru metal
by 45\% or more, the degree of 4$d$-orbital overlap would decrease
sufficiently for Ru metal to exhibit itinerant magnetism
\cite{schilling1,note1}. We note that in the itinerant ferromagnet SrRuO$_{3}%
$, the only known ferromagnetic metal among the 4$d$ oxides \cite{mazin1}, the
nearest-neighbor separation between Ru ions ($\sim$ 3.92 \AA ) is 48\% greater
than that in Ru metal (2.65 \AA ). The oxygen anions in SrRuO$_{3}$ also play
an important role in its magnetism, the Ru-O direct exchange interaction being
estimated to be strongly ferromagnetic \cite{mazin1}. Ferromagnetism would
appear to be less likely in CaRuO$_{3}$ than in SrRuO$_{3}$ since in the
former oxide the Ru-Ru nearest-neighbor separation is less and the degree of
structural disorder is greater; indeed, CaRuO$_{3}$ exhibits no magnetic order
but is paramagnetic with a negative Curie-Weiss temperature $\Theta\simeq-68$
K indicative of antiferromagnetic correlations \cite{cao1}. Both theoretical
and experimental studies have underscored the important role structural
disorder plays in suppressing the magnetic ordering temperature in these
ruthenates \cite{mazin1,cao1,dabrowski1} and other atomically arranged
perovskites \cite{dabrowski2}.

Weak itinerant ferromagnets such as ZrZn$_{2}$ \cite{pfleiderer2}, UGe$_{2}$
\cite{pfleiderer1}, MnSi \cite{pfleiderer3}, and SrRuO$_{3}$ \cite{neumeier1}
have enjoyed considerable interest for many years because of the possibility
of unconventional superconductivity, non Fermi liquid behavior, and other
exotic phenomena near quantum critical points. Relatively minor perturbations,
such as those generated by high pressures, are able to transport the system
from one ground state to another, including the destruction of the
ferromagnetic state. That the magnetism is weakened under pressure follows
from the above discussion since the degree of overlap between the magnetic
orbitals increases under pressure. Wohlfarth \cite{wohlfarth1} derived an
expression valid for weak itinerant ferromagnets whereby the rate of decrease
of the Curie temperature with pressure, $dT_{Curie}/dP,$ is inversely
proportional to the value of $T_{Curie}$ itself. It is notable that the weak
itinerant ferromagnet Sc$_{3}$In does not fit into this scheme since its Curie
temperature initially \textit{increases} under pressure \cite{grewe}.

Several years ago Belitz \textit{et al.} \cite{belitz1} argued on general
grounds that in a weakly ferromagnetic system the nature of the ferromagnetic
transition would be expected to change from second to first order at a
tricritical point, i.e. under an external perturbation such as pressure, the
order parameter (magnetization) and Curie temperature should disappear
discontinuously above a critical pressure $P_{c}.$ Metamagnetic transitions
are anticipated out of the paramagnetic state for pressures above $P_{c}$ if
strong magnetic fields are applied. Such first-order transitions under
pressure with accompanying field-induced metamagnetism have been found in the
weak itinerant ferromagnets ZrZn$_{2}$ \cite{pfleiderer2} and UGe$_{2}$
\cite{pfleiderer1}; in the latter compound the weak ferromagnetism even
coexists with superconductivity. It is clearly of interest to investigate
further systems in order to establish whether the behavior found for
ZrZn$_{2}$ and UGe$_{2}$ is only an anomaly or representative for weak
itinerant ferromagnets in general.

More than a decade ago we carried out ac susceptibility ($\chi_{ac}$) studies
in a diamond-anvil cell (DAC) on polycrystalline SrRuO$_{3}$ to nearly
hydrostatic (dense helium) pressures as high as 6 GPa \cite{neumeier1}. The
Curie temperature was found to decrease approximately linearly from 161 K to
130 K at the rate $dT_{Curie}/dP\simeq$ -5.7 K/GPa, a value somewhat less in
magnitude than that (-7.5 K/GPa) from ac susceptibility studies to 0.78 GPa by
Shikano \textit{et al.} \cite{shikano}. In two further experiments Le Marrec
\textit{et al.} \cite{marrec} and Demuer \textit{et al.} \cite{Demuer04}
carried out electrical resistivity measurements on SrRuO$_{3}$ thin films
epitaxially grown on a BaTiO$_{3}$ substrate under high quasihydrostatic
pressure (solid steatite pressure medium) to 23 and 21 GPa, respectively. In
both experiments $T_{Curie}$ was observed to decrease from 150 K at ambient
pressure to approximately 75 K at 13 GPa, yielding a negative pressure
derivative $dT_{Curie}/dP\simeq$ -5.9 K/GPa close to our value for bulk
SrRuO$_{3}$ above; however, for $T_{Curie}(P)$ was reported to pass through a
minimum near 15 GPa, increasing at higher pressures only slightly in the
former experiment \cite{marrec} but strongly in the latter \cite{Demuer04}.
This abrupt change in slope $dT_{Curie}/dP$ near 15 GPa may signal a
structural phase transition. However, the fact that the $T_{Curie}(P)$
dependences differ in the two thin-film experiments suggests that shear
stresses might play an important role; such shear stresses could arise either
from the solid steatite pressure medium or from the increasing lattice
mismatch between thin-film sample and substrate under pressure. X-ray
diffraction and magnetic susceptibility studies on bulk SrRuO$_{3}$ samples to
very high nearly hydrostatic pressures would clearly be of value to help
clarify these issues and establish the intrinsic pressure dependence
$T_{Curie}(P)$.

In the present paper we extend the pressure range of our previous ac
susceptibility measurements and include high-pressure structural studies as
well. Over the pressure range to 17.2 GPa $T_{Curie}$ is found to decrease
nearly linearly with nearly hydrostatic pressure at the rate -6.8 K/GPa with
no indication of a slope change near 15 GPa. Furthermore, ambient-temperature
X-ray diffraction studies to 25.3 GPa reveal that SrRuO$_{3}$ remains in the
orthorhombic structure over the entire pressure range. Interestingly, the
pressure dependences of the lattice parameters are quite complex but allow the
estimate that the average Ru-O-Ru bond angle $\beta$ initially decreases under
pressure but passes through a minimum near 15 GPa. A possible correlation
between the pressure dependence of the Curie temperature $T_{Curie}$ and the
bond angle $\beta$ is discussed. The present results make it imperative that
the magnetic and superconducting properties of this interesting system be
extended to even higher pressures and lower temperatures.

\section{Experiment}

The polycrystalline sample of SrRuO$_{3}$ used in the present experiments was
prepared by solid state reaction of stoichiometric quantities of high purity
(99.9\% or better) SrCO$_{3}$ and RuO$_{2}$. The starting materials were
weighed, mixed and placed in an Al$_{2}$O$_{3}$ crucible. The specimen was
reacted for 4 hours at 1100$^{\circ}$C, ground with an agate mortar and pestle
for 5 min and reacted in air for 6 h at 1250$^{\circ}$C. Subsequently, the
specimen was reground for 5 min and reacted at 1250$^{\circ}$C for 14 h and
cooled by shutting the furnace off. Powder X-ray diffraction proved the
specimen to be single-phase with no observable secondary phases.

Single-crystals of SrRuO$_{3}$ were grown in Pt crucibles from
off-stoichiometric quantities of RuO$_{2}$, SrCO$_{3}$, and SrCl$_{3}$
mixtures with SrCl$_{2}$ self flux. The mixtures were first heated to
1500$^{\circ}$C, soaked for 25 hours, slowly cooled at 2-3$^{\circ}$C/hour to
1350$^{\circ}$C, and finally cooled to room temperature at 100$^{\circ}%
$C/hour. The single crystals were characterized by single crystal X-ray
diffraction and by scanning and transmission electron microscopies. All
results indicate that the single crystals are of high quality.

High pressure X-ray diffraction experiments were performed in a DAC on
polycrystalline SrRuO$_{3}$ using synchrotron X-rays ($\lambda=0.41105$ \AA )
with beam size $20\times20$ $\mu$m$^{2}$ at the Advanced Photon Source HPCAT,
Sector 16-IDB, Argonne National Labs. The powder sample was loaded together
with a ruby chip and silicon fluid as pressure medium into a 135 $\mu$m
diameter bore through a rhenium gasket \cite{shen1}. The X-ray diffraction
patterns (see Fig.~1) were collected using a MAR image plate camera with
$100\times100$ $\mu$m$^{2}$ pixel dimension for 10-20 s. The images were
integrated using FIT2D program \cite{hammersley1} and structural refinements
were carried out using JADE \cite{jade1}.

High-pressure ac magnetic susceptibility measurements $\chi_{ac}(T)$ were
carried out using a DAC made of CuBe alloy \cite{schilling1} where two
1/6-carat type Ia diamond anvils with 0.5 mm diameter culets press onto a 3 mm
diameter gold-plated rhenium gasket preindented from 250 $\mu$m to 80 $\mu$m
thickness and containing a centered 235 $\mu$m diameter hole. After the sample
(typical dimensions $80\times80\times30$ $\mu$m$^{3})$ and tiny ruby spheres
\cite{chervin} are placed in the hole, the DAC is assembled, cooled down to
low temperatures and flooded with liquid helium pumped to 2 K. The opposing
diamond anvils are then pressed into the gasket to trap and build up pressure
in the liquid helium pressure medium surrounding the sample. As in the above
X-ray studies, the standard ruby calibration \cite{mao1} is used to determine
the pressure to within $\pm0.2$ GPa. The temperature was kept below 180 K
during the entire experiment to reduce the chance that helium might enter the
diamond anvils, possibly causing them to fail.

The ferromagnetic transition at $T_{Curie}$ is determined inductively using
two balanced primary/secondary coil systems located immediately outside the
metal gasket \cite{tomita1} and connected to a Stanford Research SR830 digital
lock-in amplifier. The ac susceptibility studies were carried out using a 3 Oe
r.m.s. magnetic field at 1023 Hz. Further experimental details of the
diamond-anvil cell and ac susceptibility techniques are published elsewhere
\cite{schilling1,hamlin1}.

The lock-in amplifier allows the measurement of not only the basic 1st
harmonic of the ac susceptibility, $\chi_{1}$, but also the higher harmonics.
Since in this experiment the signal from the ferromagnetic transition of
SrRuO$_{3}$ in $\chi_{1}$ becomes very difficult to resolve for pressures
above 10 GPa, we decided to also measure the 3rd harmonic $\chi_{3}$ which has
a superior signal/noise ratio at the highest pressures. In contrast to
$\chi_{1}(T),$ where the transition is only revealed after the subtraction of
a relatively large temperature-dependent background, the background signal in
$\chi_{3}(T)$ has little temperature dependence. In Fig.~2 we compare at
ambient pressure the real part of the 1st harmonic $\chi_{1}^{\prime}(T)$ to
the imaginary part of the 3rd harmonic $\chi_{3}^{\prime\prime}(T)$ for a
single-crystalline sample much larger ($150\times250\times65$ $\mu$m$^{3}$)
than that used in a typical high-pressure experiment in the DAC. It is seen
that the temperature of the peak in $\chi_{1}^{\prime}(T)$, which we use to
define the Curie temperature $T_{Curie},$ corresponds well with the transition
onset in $\chi_{3}^{\prime\prime}(T).$ As we will see, measuring the 3rd
harmonic allows us to follow the ferromagnetic transition in SrRuO$_{3}$ to
higher pressures.

\section{Results of Experiment}

\subsection{X-ray Diffraction Studies}

At ambient pressure SrRuO$_{3}$ is found to crystallize in the orthorhombic
structure (Pbnm, 62) with lattice parameters $a=5.5754(7)$ \AA , $b=5.5405(6)
$ \AA , $c=7.8546(8)$ \AA , in good agreement with literature values
\cite{cao1,Kobayashi94}. Fig.~1 shows the X-ray diffraction patterns at
various pressures to 25.3 GPa. As pressure is increased, no new peaks are
found indicating the sample retains the orthorhombic crystal structure.
However, the peaks broaden due to increasing nonhydrostatic conditions in the
solidified silicone pressure medium. The pressure dependences of the lattice
parameters are given in Table I and are shown in Fig.~3 (upper). The smooth
monotonic change in all lattice parameters under pressure to 25.3 GPa is
consistent with the absence of even an isostructural phase transition over
this pressure range. However, it is seen that there is a crossover of the $b$
and $c/\sqrt{2}$ parameters near 15 GPa; this will be discussed in detail below.

The equation of state for SrRuO$_{3}$ is shown in Fig.~3 (lower) with the
least-squares fit (solid line) obtained using the Birch-Murnaghan expression
\cite{birch1}%

\begin{equation}
P=\frac{3}{2}B_{0}\left[  \left(  V/V_{0}\right)  ^{-7/3}-\left(
V/V_{0}\right)  ^{-5/3}\right]  \times\left\{  1+\frac{3}{4}\left(
B_{0}^{\prime}-4\right)  \left[  \left(  V/V_{0}\right)  ^{-2/3}-1\right]
\right\}  ,
\end{equation}
where $B_{0}=192(3)$ GPa is the bulk modulus and $B_{0}^{\prime}=5.0(3)$ its
pressure derivative. These values are similar to those found in Ca-based
perovskites with Pbnm symmetry and similar ambient pressure volumes
\cite{Xiong86,Ross99,Kung01}.

Again, we find no evidence for a structural phase transition near $P\approx15$
GPa where abrupt changes in $dT_{Curie}/dP$ were seen for two different
thin-film SrRuO$_{3}$ samples \cite{marrec,Demuer04}. However, this pressure
lies near that for the lattice parameter crossover seen in Fig.~3 (upper). As
it is known that $T_{Curie}$ is reduced \cite{cao1,shikano,Kobayashi94} as one
moves farther away from cubic symmetry by substituting Ca for Sr, it is
instructive to ask whether pressure might have a similar effect. Following the
work of others \cite{Okeeffe77,Zhao93}, if one assumes that the RuO$_{6}$
coordination octahedra are not distorted and only rotate when pressure is
applied (a reasonable assumption for the small rotations occurring in
SrRuO$_{3}$), the angles of rotation around the characteristic directions of
the ideal cubic perovskite structure can be calculated by simply using the
orthorhombic lattice parameters. The validity of using this model in a
semi-quantitative manner for determining the tilt angles has been shown by the
good agreement between the angles estimated from the lattice parameters and
directly from the atomic positions \cite{Lin05}. The Ru-O-Ru bond angles can
be found simply from the angle of rotation $\Phi$ around the (111)$_{p}$
direction of the ideal cubic perovskite structure given by
\begin{equation}
\Phi=\cos^{-1}\left(  \frac{\sqrt{2}b^{2}}{ac}\right)  .
\end{equation}
The Ru-O-Ru bond angles $\beta_{1}$ and $\beta_{2}$ are then given by%
\begin{align}
\beta_{1}  &  =\cos^{-1}\left(  \frac{2-5\cos^{2}\Phi}{2+\cos^{2}\Phi}\right)
\text{ and}\\
\beta_{2}  &  =\cos^{-1}\left(  \frac{1-4\cos^{2}\Phi}{3}\right)  .\nonumber
\end{align}
While the absolute values of the angles may not be in exact agreement, the
relative change of the angles obtained using these formula agrees well with
high resolution diffraction data. In fact, as one goes from SrRuO$_{3}$ to
CaRuO$_{3}$, direct calculation of the average Ru-O-Ru bond angle $\beta$
decreases by 14.1$^{\circ}$ \cite{Kobayashi94}, while the estimate from the
lattice parameters is 15.4$^{\circ}$. The inset in the bottom panel of Fig.~3
shows the pressure dependence of the average Ru-O-Ru bond angle calculated
from the lattice parameters. Up to approximately 15 GPa we find that the value
of the angle $\beta$ decreases in a nearly linear fashion ($d\beta
/dP\simeq-0.86$ $\deg$/GPa) signifying that the perovskite structure is
becoming more distorted. However, for pressures above 15 GPa the angle $\beta$
is seen to increase! We explore the significance of this finding below.

We have also found that these structural results are reproduced when Ar or a
4:1 methanol:ethanol mixture is used as the pressure transmitting media
instead of silicon oil. This leads us to the conclusion that the results of
our structural studies on bulk samples are not unduly sensitive to the
pressure transmitting fluid used. This along with detailed structural
information will be presented in a future report \cite{Kumar07}.

\subsection{ac Susceptibility Studies}

The first of the present experiments (run B) was carried out on a
polycrystalline SrRuO$_{3}$ sample (dimensions $110\times130\times30$ $\mu
$m$^{3}$) in a DAC under nearly hydrostatic pressures to 14.7 GPa. In Fig.~4
the results are compared to those from our earlier measurements to 6 GPa (run
A) \cite{neumeier1}; the agreement is reasonable, although the final two data
points in the earlier study lie somewhat above the straight-line fit. From 0
to 14.7 GPa the height of the transition in the ac susceptibility decreased
from approximately 5 to 1.5 nV. One should not conclude from this that the
value of the magnetic moment per ion necessarily decreases since the ac
susceptibility measures only the initial response of the sample magnetization
to an imposed magnetic field and thus also depends on extrinsic parameters
such as the degree of domain wall pinning. A pressure-dependent reduction in
the applied field by shielding currents in the Re gasket may also contribute
to the signal reduction. The failure of the diamond anvils ended the
experiment in run B.

Run B was followed by run C on a polycrystalline sample (dimensions
$68\times90\times25$ $\mu$m$^{3}$) to 9 GPa; as seen in Fig.~4, the results
are in good agreement with those from run B. From 0 to 9 GPa the height of the
transition decreased from 2.5 to 0.6 nV. At the next higher pressure (15 GPa)
the transition had decreased sufficiently in size that it was not possible to
unequivocally identify it.

The next high-pressure DAC experiment (run D) was carried out on a
single-crystalline SrRuO$_{3}$ sample (dimensions $65\times90\times30$ $\mu
$m$^{3}$) where both $\chi_{1}^{\prime}(T)$ and $\chi_{3}^{\prime\prime}(T)$
were measured in an effort to track the ferromagnetic transition to higher
pressures (Fig.~5). The transition height in $\chi_{1}^{\prime}(T)$ is seen to
decrease roughly by a factor of two from 0 to 9.9 GPa, but the transition also
becomes broader, perhaps due to nonhydrostatic stresses even in the helium
pressure medium; the area under the transition curve decreases by only $\sim$
20\% over the pressure range to 9.9 GPa. At higher pressures the transition
could not be identified in the 1st harmonic, so $\chi_{3}^{\prime\prime}(T)$
was measured (see the lower half of Fig.~5). Even though the transition in
$\chi_{3}^{\prime\prime}(T)$ is relatively small, the absence of a strong
temperature-dependent background signal allows one to track $T_{Curie}(P)$ to
higher pressures. The dependence of $T_{Curie}$ on pressure is seen in Fig.~4
to be reversible and approximately linear to 17.2 GPa with slope
$dT_{Curie}/dP\simeq-6.8$ K/GPa, $T_{Curie}$ decreasing from 162 K to 42.7 K,
i.e. by nearly a factor of four. Unfortunately, at higher pressures (20.4 to
34 GPa) the ferromagnetic transition could no longer be resolved.

\section{Conclusions}

As mentioned in the Introduction, in a canonical weak itinerant ferromagnet
the Curie temperature and saturation magnetization would be expected to
decrease monotonically under pressure, falling to zero at a critical pressure
$P_{c}$ \cite{wohlfarth1}. That SrRuO$_{3}$ may exhibit more complex behavior
is indicated by the fact that in thin-film studies \cite{marrec,Demuer04} the
Curie temperature passes through a minimum near 15 GPa (Fig.~4), approximately
the same pressure where we find the average Ru-O-Ru bond angle $\beta$ in a
crystalline sample to pass through a minimum (Fig.~3). Indeed, in Sr$_{1-x}%
$Ca$_{x}$RuO$_{3}$ and related systems the value of $T_{Curie}$ appears to be
inversely related to the degree of distortions away from the ideal cubic
perovskite structure, i.e. the greater the distortions, the lesser the average
bond angle $\beta$ and the lower the value of $T_{Curie}$
\cite{mazin1,cao1,dabrowski1}. We can estimate the dependence of $T_{Curie}$
on $\beta$ by considering that upon increasing $x$ in Sr$_{1-x}$Ca$_{x}%
$RuO$_{3}$ from 0 to 1 the value of $T_{Curie}$ decreases from 162 K to 0 K
whereas the bond angle $\beta$ decreases by 15.4$^{\circ}$ \cite{cao1},
yielding $dT_{Curie}/d\beta\approx+10.5$ K/$\deg. $ We note that this value of
$dT_{Curie}/d\beta$ is quite close to that derived from the values of
$dT_{Curie}/dP$ and $d\beta/dP$ in the present high-pressure experiments from
0 to 15 GPa where $dT_{Curie}/d\beta=(dT_{Curie}/dP)/(d\beta/dP)\approx(-6.8$
K/GPa$)/(-0.86$ $\deg$/GPa$)=+7.9$ K/$\deg.$ The present experiments on the
single-crystalline SrRuO$_{3}$ sample thus support the contention that
$T_{Curie}$ is correlated with the degree of structural distortion as
represented by the bond angle $\beta, $ at least in the pressure range 0 - 15 GPa.

For pressures at 15 GPa and above, $\beta$ passes through a minimum and then
increases quite rapidly, as seen in Fig.~3. The above correlation between
$T_{Curie}$ and $\beta$ for bulk SrRuO$_{3}$ would then lead to the
expectation that $T_{Curie}(P)$ should also pass through a minimum near 15 GPa
and increase at higher pressures. Unfortunately, this possibility cannot be
adequately checked in the present experiment since the anticipated minimum in
$T_{Curie}(P)$ at 15 GPa lies close to the maximum pressure (17.2 GPa) for
which the ferromagnetic transition can be resolved.

It is not unreasonable to assume that $T_{Curie}(P)$ may pass through a
minimum at lower pressures in thin-film relative to bulk samples since the
former are already in a strained state at ambient pressure which is consistent
with their lower ambient pressure value of $T_{Curie}$. In the thin-film
studies shear stresses may originate either from the solid pressure medium or
from the increasing lattice mismatch between the thin-film SrRuO$_{3}$ sample
and its CaTiO$_{3}$ substrate; the latter effect is expected since the bulk
moduli $B_{o}$ of SrRuO$_{3}$ (192(3) GPa) and CaTiO$_{3}$ (176 GPa from
Ref.~\cite{edwards}) differ by 9\%. The fact that for both thin film samples a
break in slope in $T_{Curie}(P)$ is observed near 15 GPa with quite different
behavior at higher pressures can easily be explained by differing strain
states in the two thin film samples. This would lead to different values of
the Ru-O-Ru angles, and hence different $T_{Curie}$ values, as a function of pressure.

A search for superconductivity in single-crystalline SrRuO$_{3}$ with negative
results was carried out by measuring $\chi_{1}^{\prime}(T)$ or $\chi
_{3}^{\prime\prime}(T)$ to temperatures as low as 4 K at the following
pressures: \ 17.2, 20.4, 21.3, 22.5, 23.8, 25.8, and 34 GPa. Similar ac
susceptibility studies on polycrystalline CaRuO$_{3}$ at 6 and 8 GPa found no
evidence for either ferromagnetism or superconductivity above 4 K.

In summary, parallel ac susceptibility and X-ray diffraction studies have been
carried out on poly- and single-crystalline samples of SrRuO$_{3}$ to 34 and
25.3 GPa, respectively. The structural studies yield the unexpected result
that the lattice distortions increase with pressure to a maximum value near 15
GPa but then decrease (the bond angle $\beta$ passes through a minimum near 15
GPa). $T_{Curie}(P)$ is found to decrease nearly linearly to 17.2 GPa nearly
hydrostatic pressure. Evidence is given for a possible correlation between
$T_{Curie}(P)$ and $\beta(P)$ in the present experiments which would appear to
imply that $T_{Curie}(P)$ should also pass through a minimum near 15 GPa,
contrary to the expectations for a canonical weak itinerant ferromagnet where
$T_{Curie}$ would fall monotonically to 0 K under pressure. Future experiments
on bulk SrRuO$_{3}$ to even higher pressures and lower temperatures are
imperative to clarify the magnetic and superconducting properties of this
interesting system.\vspace{0.5cm}

\noindent\textbf{Acknowledgments. }This material is based upon work supported
by the National Science Foundation through grants DMR-0404505 (Washington
University) and DMR-0504769 (Montana State University). Work at UNLV is
supported by DOE, NNSA, under Cooperative Agreement DE-FC08-01NV14049.

\newpage\begin{table}[ptb]
\caption{High pressure structural parameters for SrRuO$_{3}$. The lattice
parameters are for the orthorhombic Pbnm (\#62) structure. The values in
parentheses represent the estimated uncertainty in pressure and the error in
least significant digits from the standard errors in unit cell
refinements.\vspace{0.6cm}}%
\label{latticeparams}
\begin{tabular}
[c]{ccccc}\hline\hline
$P$ (GPa) & $a$ (\AA ) & $b$ (\AA ) & $c$ (\AA ) & $V$ (\AA )\\\hline
$0$ & $5.5754(7)$ & $5.5405(6)$ & $7.8546(8)$ & $242.62(5)$\\
$0.6(2)$ & $5.561(1)$ & $5.530(2)$ & $7.851(2)$ & $241.43(11)$\\
$2.8(2)$ & $5.549(1)$ & $5.505(1)$ & $7.825(2)$ & $239.05(9)$\\
$3.3(2)$ & $5.550(1)$ & $5.503(1)$ & $7.808(2)$ & $238.46(9)$\\
$5.2(2)$ & $5.531(1)$ & $5.480(2)$ & $7.791(2)$ & $236.15(11)$\\
$8.5(3)$ & $5.529(2)$ & $5.439(2)$ & $7.741(3)$ & $232.81(13)$\\
$10.0(4)$ & $5.517(2)$ & $5.419(2)$ & $7.747(4)$ & $231.58(17)$\\
$11.6(4)$ & $5.530(2)$ & $5.413(2)$ & $7.686(3)$ & $230.07(15)$\\
$12.8(5)$ & $5.507(2)$ & $5.412(2)$ & $7.685(3)$ & $229.03(15)$\\
$14.1(5)$ & $5.503(2)$ & $5.538(2)$ & $7.700(3)$ & $227.98(15)$\\
$17.5(5)$ & $5.515(2)$ & $5.380(2)$ & $7.594(3)$ & $225.31(15)$\\
$18.6(5)$ & $5.510(3)$ & $5.391(2)$ & $7.551(3)$ & $224.30(17)$\\
$20.1(5)$ & $5.499(1)$ & $5.416(1)$ & $7.488(2)$ & $223.00(8)$\\
$21.9(5)$ & $5.486(2)$ & $5.394(2)$ & $7.493(6)$ & $221.73(21)$\\
$23.6(5)$ & $5.480(1)$ & $5.391(1)$ & $7.472(2)$ & $220.74(8)$\\
$25.3(5)$ & $5.470(1)$ & $5.370(1)$ & $7.471(1)$ & $219.49(6)$\\\hline\hline
\end{tabular}
\end{table}

\begin{center}
\newpage\bigskip{\LARGE Figure Captions}
\end{center}

\bigskip\ 

\noindent\textbf{Fig. 1. \ }X-ray diffraction pattern from a powdered
polycrystalline SrRuO$_{3}$ sample at ambient temperature for various
pressures from\ 2.8 to 25.3 GPa. Over this pressure range the crystal
structure remains orthorhombic.\bigskip

\noindent\textbf{Fig. 2. \ }Temperature dependence of the ac susceptibility
$\chi(T)$ of single-crystalline SrRuO$_{3}$ at ambient pressure. $\chi
_{1}^{\prime}$ gives the real part of the 1st harmonic and $\chi_{3}%
^{\prime\prime}$\ gives the imaginary part of the 3rd harmonic in nanovolt
units; the scale of $\chi_{3}^{\prime\prime}$ is expanded 20$\times.$ The
Curie temperature $T_{Curie}\simeq162$ K is defined by the peak in $\chi
_{1}^{\prime}(T)$ or, equivalently, by the onset in $\chi_{3}^{\prime\prime
}(T),$ as shown in the figure.\bigskip

\noindent\textbf{Fig. 3. \ }(upper) Lattice parameters of orthorhombic
SrRuO$_{3}$ versus pressure to 25.3 GPa at ambient temperature. Solid lines
are guides to eye. (lower) Equation of state from data in upper figure. Data
fit using Eq.~1 (solid line) yields the bulk modulus $B_{o}=192(3)$ GPa and
its pressure derivative $B_{o}^{\prime}=5.0(3).$ Inset shows average Ru-O-Ru
bond angle versus pressure (see text).\bigskip

\noindent\textbf{Fig. 4. \ }Dependence of the Curie temperature $T_{Curie}$ of
SrRuO$_{3}$ on nearly hydrostatic pressure to 17.2 GPa. $T_{Curie}$ is seen to
decrease monotonically with pressure in reversible fashion. Thick solid
straight line gives fit to present data with slope $dT_{Curie}/dP\simeq-6.8$
K/GPa. Numbers give order of present measurements (run D) on single crystal
($\bullet$ from $\chi_{1}^{\prime}$, $\blacksquare$ from $\chi_{3}%
^{\prime\prime}$); also shown are present results on polycrystals ($\Diamond$
run B) and ($\bigtriangleup$ run C) as well as previous results
($\bigtriangledown$ run A) from Ref.~\cite{neumeier1}. Thin solid and dashed
lines give results from thin-film studies from Refs.~\cite{Demuer04} and
\cite{marrec}, respectively.\bigskip

\noindent\textbf{Fig. 5. \ }Temperature dependence of the ac susceptibility of
single-crystalline SrRuO$_{3}$ at nearly hydrostatic pressures to 17.2 GPa
(run D in text). Above and below the horizontal dotted line are shown,
respectively, the measured ferromagnetic transitions in $\chi_{1}^{\prime}(T)$
and $\chi_{3}^{\prime\prime}(T)$ from which $T_{Curie}$ is determined (see
Fig.~2). The numbers give the order of measurements corresponding to the
single-crystal data ($\bullet,\blacksquare$) in Fig.~4.


\begin{thebibliography}{99}
\bibitem {note1}In elemental solids the degree of overlap between neighboring
partially filled orbitals determines whether or not the solid is magnetic. We
define $\alpha\equiv r_{a}/r_{m},$ where $r_{m}$ is the radius of maximum
charge density for the magnetic orbital and $r_{a}=V_{a}^{1/3}/2$ is half the
average interatomic separation, $V_{a}$ being the available volume per atom.
$\alpha$ is thus inversely related to the degree of orbital overlap. It has
been shown that for $\alpha>\alpha_{c}=3.2$ the degree of overlap is
sufficiently small that the system is magnetic, but nonmagnetic for
$\alpha<\alpha_{c}=3.2$ \cite{schilling1}. Indeed, it is found that magnetism
prevails in elemental solids from all rare-earth elements (4$f$), as well as
from the heavy 3$d$ transition-metal elements and heavy actinides (5$f$) where
in all cases $\alpha>3.2$. On the other hand, for all the 4d and 5d
transition-metal elements $\alpha\approx$ 1.9 to \ 2.4 which is less than
$\alpha_{c}=3.2$. This critical value exceeds that for Ru metal ($\alpha
\simeq$ 2.2) by 45\%, so that the lack of magnetic behavior in Ru metal comes
as no surprise.
\end{thebibliography}
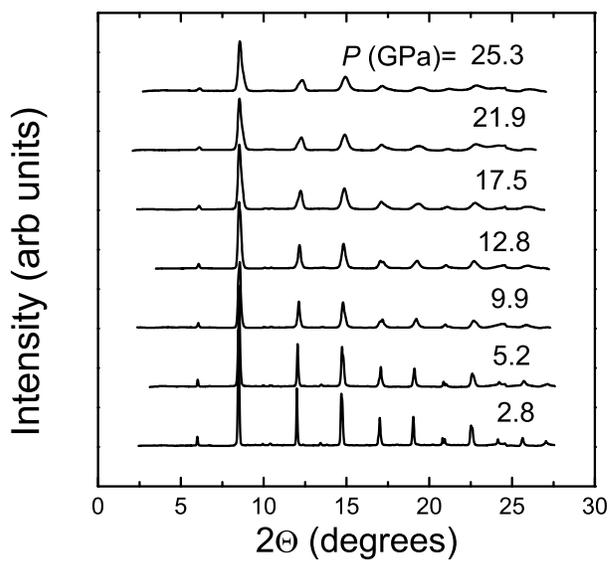
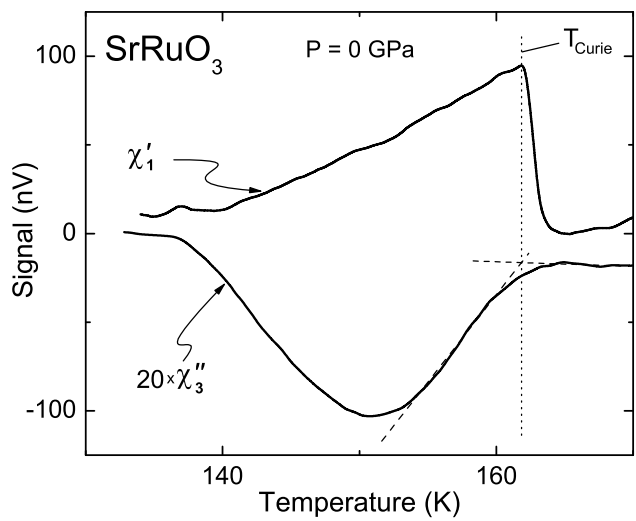
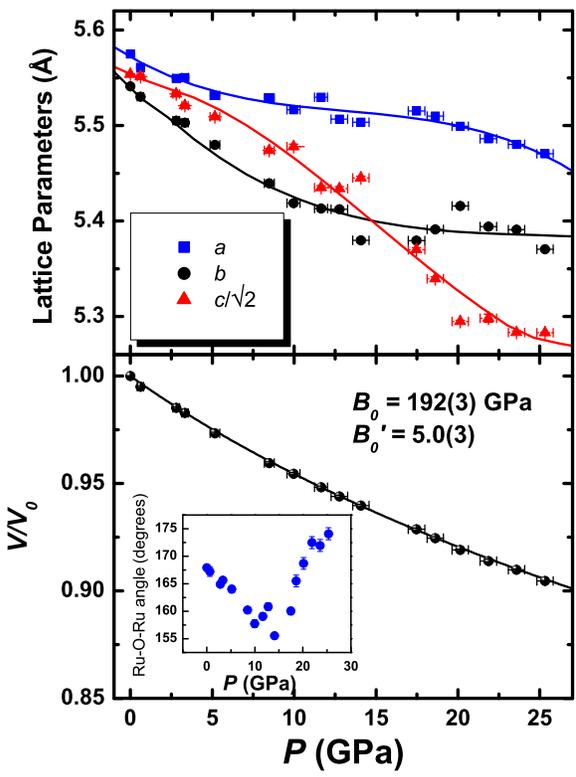
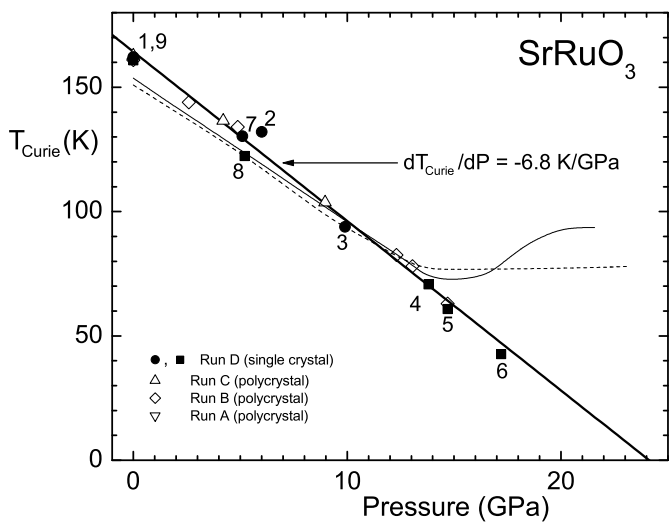
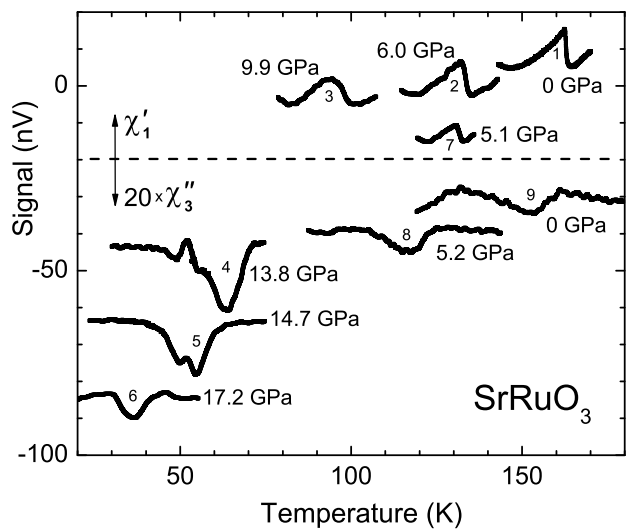
\end{document}